\RequirePackage[table]{xcolor}%

\documentclass[AMA,STIX1COL]{WileyNJD-v2}

\usepackage{siunitx}
\usepackage{hyperref}
\usepackage{url}

\newcommand\blfootnote[1]{%
  \begingroup
  \renewcommand\thefootnote{}\footnote{#1}%
  \addtocounter{footnote}{-1}%
  \endgroup
}

\usepackage{adjustbox}
\usepackage{array}
\usepackage{multirow}
\usepackage{booktabs}

\usepackage{colortbl}
\usepackage{xspace}

\usepackage{lscape}
\usepackage{longtable}

\usepackage{amsmath}
\usepackage{graphicx}
\usepackage[colorinlistoftodos]{todonotes}
\usepackage{hyperref}
\usepackage{url}

\articletype{Special issue}%

\received{27 July 2018}
\revised{27 July 2019}
\accepted{12 September 2019}

\newcommand{\uib}[1]{#1\xspace}

\begin{document}

\title{How to Place Your Apps in the Fog\\ {\Large State of the Art and Open Challenges}}

\author[1]{Antonio Brogi}

\author[1]{Stefano Forti}

\author[2]{Carlos Guerrero}

\author[2]{Isaac Lera}

\authormark{BROGI, FORTI, GUERRERO and LERA}

\address[1]{\orgdiv{Department of Computer Science}, \orgname{University of Pisa}, \orgaddress{\state{Pisa}, \country{Italy}}}

\address[2]{\orgdiv{Department of Computer Science}, \orgname{University of the Balearic Islands}, \orgaddress{\state{Balearic Islands}, \country{Spain}}}

%\address[3]{\orgdiv{Org Division}, \orgname{Org Name}, \orgaddress{\state{State name}, \country{Country name}}}

\corres{Stefano Forti, \email{stefano.forti@di.unipi.it}}

\fundingInfo{
University of Pisa project ``\textit{DECLWARE: Declarative methodologies of application design and deployment}" (PRA\_2018\_66) and by the Spanish Government (Agencia Estatal de Investigaci\'on) and the European Commission (Fondo Europeo de Desarrollo Regional), Grant Number: TIN2017-88547-P (MINECO/AEI/FEDER, UE)\\
\\
\textbf{Authors Contribution:} The authors are ordered alphabetically. All the authors contributed equally to this work.
}

\abstract[Summary]{Fog computing aims at extending the Cloud towards the IoT so to achieve improved QoS and to empower latency-sensitive and bandwidth-hungry applications. The Fog calls for novel models and algorithms to distribute multi-service applications in such a way that data processing occurs wherever it is best-placed, based on both functional and non-functional requirements.

\noindent This survey reviews the existing methodologies to solve the application placement problem in the Fog, while pursuing three main objectives. First, it offers a comprehensive overview on the currently employed algorithms, on the availability of open-source prototypes, and on the size of test use cases. Second, it classifies the literature based on the application and Fog infrastructure characteristics that are captured by available models, with a focus on the considered constraints and the optimised metrics. Finally, it identifies some open challenges in application placement in the Fog.}


\jnlcitation{\cname{%
\author{A. Brogi}, 
\author{S. Forti}, 
\author{C. Guerrero}, and 
\author{I. Lera}} (\cyear{2018}), 
\ctitle{Place Your Apps in the Fog. State of the Art and Open Challenges}, \cjournal{Software Practice Experience}, \cvol{2019;1--22}, \url{https://doi.org/10.1002/spe.2766}.}

\maketitle

%\footnotetext{\textbf{Abbreviations:} FAPP - Fog Application Placement Problem.}
\blfootnote{{\footnotesize \textcopyright\ 2019. This manuscript version is made available under the CC-BY-NC-ND 4.0 license  (\url{http://creativecommons.org/licenses/by-nc-nd/4.0/}). This manuscript is a preprint of the article published in  {\it Software Practice and Experience} and available at: \url{https://doi.org/10.1002/spe.2766}}}

%\documentclass[a4paper,twocolumn]{article}

%% Language and font encodings
%\usepackage[english]{babel}
%\usepackage[utf8x]{inputenc}
%\usepackage[T1]{fontenc}

%rotate column titles
%\usepackage{adjustbox}
%\usepackage{array}

%\newcolumntype{R}[2]{%
%    >{\adjustbox{angle=#1,lap=\width-(#2)}\bgroup}%
%    l%
%    <{\egroup}%
%}
%\newcommand*\rot{\multicolumn{1}{R{90}{1em}}}% no optional argument here, please!
%%%

%\usepackage{lscape}

%% Sets page size and margins
%\usepackage[a4paper,top=3cm,bottom=2cm,left=3cm,right=3cm,marginparwidth=1.75cm]{geometry}

%% Useful packages
%\usepackage{amsmath}
%\usepackage{graphicx}
%\usepackage[colorinlistoftodos]{todonotes}
%\usepackage[colorlinks=true, allcolors=blue]{hyperref}

%\title{How to Place Your Fog Applications\\ {\large State of the Art and Open Challenges}}
%\author{UIB - UPI}

%\begin{document}
%\maketitle

%\begin{abstract}

%\end{abstract}

\section{Introduction}
\label{intro}
  
CISCO expects more than 50 billion of connected entities (people, machines and connected Things) by 2021, and estimates they will have generated around 850 Zettabytes of information by that time, of which only 10$\%$ will be useful to some purpose~\cite{cisco, cisco2}. As a consequence of this trend, enormous amounts of data -- the so-called \textit{Big Data}~\cite{bigdata} -- are collected by Internet of Things (IoT) sensors and stored in Cloud data centres~\cite{bigdataissues}. 
Once there, data are subsequently analysed to determine reactions to events or to extract analytics or statistics. Whilst data-processing speeds have increased rapidly, bandwidth to carry data to and from data centres has not increased equally fast~\cite{dustdar}. On one hand, supporting the transfer of data from/to billions of IoT devices is becoming hard to accomplish due to the volume and geo-distribution of those devices. On the other hand, the need to reduce latency for time-sensitive applications, to eliminate mandatory connectivity requirements, and to support computation or storage closer to where data is generated 24/7, is evident~\cite{buyyacomputer}. 

\begin{figure}[bt]
\centering
\includegraphics[width=10cm]{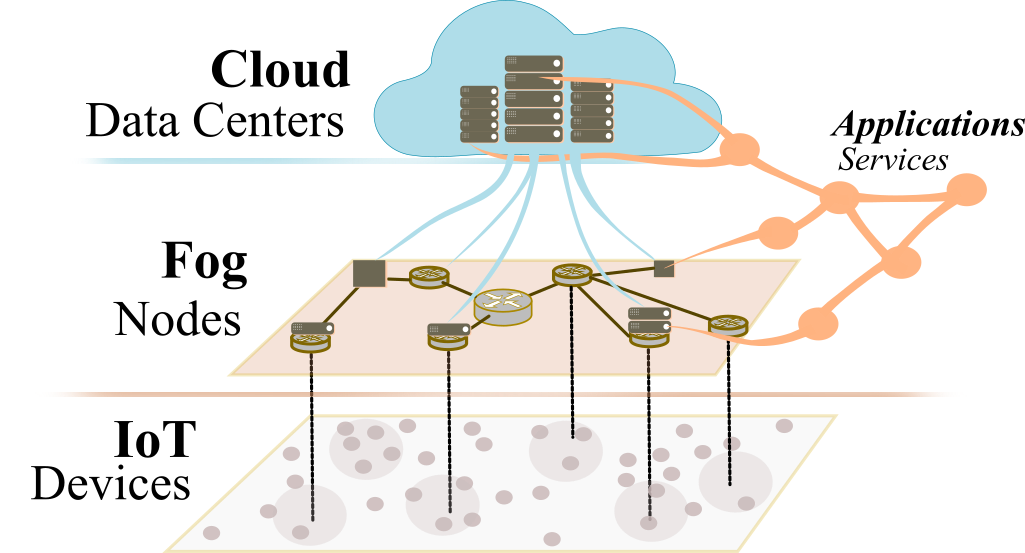}
\caption{\uib{Fog computing architecture.}}
\label{layerFogFig}
\end{figure}

\medskip
In this context, a new utility computing paradigm took off, aiming at connecting the ground (IoT) to the sky (Cloud), and it has been named \textit{Fog computing}~\cite{bonomifog}. The Fog aims at better supporting time-sensitive and bandwidth hungry IoT applications by selectively pushing \uib{their components/services} closer to where data is produced and by exploiting a geographically distributed multitude of heterogeneous devices (e.g.,  personal devices, gateways, micro-data centres, embedded servers) spanning the continuum from the IoT to the Cloud \uib{(Fig.~\ref{layerFogFig})}. %\uib{ shows three conceptual layers on current network infrastructure, where applications or services can be allocated in Cloud data centers or intermediate network devices.} %A substantial amount of computation, storage and networking is therefore expected to happen closer to where data is produced and to cyber-physical systems based on the IoT, contiguously to and interdependently with the Cloud. 
In its reference architecture\footnote{The OFC reference architecture of Fog computing was also adopted as the IEEE 1934-2018 standard \cite{ieee1934}.}, the OpenFog Consortium (OFC) \cite{openfogarch}, which is fostering academic and industrial research in the field since 2015, gives the following definition of Fog computing: 

\begin{center}
\it
Fog computing is a system-level horizontal architecture that distributes resources and services of computing, storage, control and networking anywhere along the continuum from Cloud to Things, thereby accelerating the velocity of decision making.  Fog-centric architecture serves a specific subset of business problems that cannot be successfully implemented using only traditional cloud-based architectures or solely intelligent edge devices.
\end{center}

%TODO DOUBLE-CHEK: Design?, reference in text?, caption?

\noindent The NIST  has also recently proposed a conceptual architecture for Fog computing~\cite{nistconceptual}. The Fog configures as a powerful enabling complement to the IoT+Edge and to the IoT+Cloud scenarios, featuring a new intermediate layer of cooperating devices that can autonomously run services and complete specific business missions, contiguously with the Cloud and with cyber-physical systems at the edge of the network~\cite{brogi2018bonsai}.

\medskip
Overall, Fog computing should ensure that computation over the collected data happens wherever it is \textit{best-placed}, based on various application (e.g., hardware, software, QoS) or stakeholders (e.g., cost, business-related) requirements. Since modern software systems are more and more often made from distributed, (numerous) interacting components (e.g., service-oriented and micro-service based architectures), it is challenging to determine where to deploy each of them so to fulfil all set requirements. 
\uib{Fog computing architectures substantially differ from Cloud architectures. Particularly, according to recent literature~\cite{moysiadis2018,DBLP:conf/fwc/MayerGSR17,doi:10.1002/cpe.5343,garcia2015edge,fmec19}: 
\begin{itemize}
    \item[--] Fog nodes feature limited and very heterogeneous resources, whilst data centre nodes feature high (and virtually unbounded) computational, storage and power capabilities,
    \item[--] Fog devices are highly geographically distributed -- often mobile -- and possibly span wide large-scale networks reaching closer to (human and machine) end-users, whilst Cloud data centres are located in few geographic locations all over the world and connect directly to fibre backbones,    
    \item[--] end-to-end latency between Cloud nodes within data centres is usually negligible and bandwidth availability is guaranteed via redundant links, whilst in Fog domains network QoS between nodes can largely vary due to the presence of a plethora of different (wired or wireless) communication and Internet access technologies,
    %\item[(d)] nodes in data centers are usually connected via redundant links, in contrast to Fog networks, where regions of the network can be connected to the rest of the network with a single link, which, for example, could be a wireless one; and 
    %\item[(d)] fog devices are very heterogeneous devices, in contrast to data center devices, which usually have the same characteristics or very similar ones.  
    \item[--] Fog nodes are owned and managed by various service providers (from end-users to Internet Service Providers to  Cloud operators) and might also opportunistically include  available edge devices (e.g., crowd-computing, \textit{ad-hoc} networks) whereas largest Cloud data centres are in the hands of a few big players.
\end{itemize}
}

\noindent 
\uib{
Due to these differences, Fog computing calls for new and specific methodologies to optimise the distribution of application functionalities and the usage of the available infrastructure, by suitably dealing with (possibly) limited hardware resources and large-scale deployments, unstable connectivity and platform/operators heterogeneity, and (node or link) failures \cite{mann2019optimization}.
%These differences make it necessary to re-evaluate, or define new ones, traditional optimization policies.
 }
Lately, following this line, a significant amount of research has considered the problem of \textit{optimally} placing application components or services based on different -- and sometimes orthogonal -- constraints. However, to the best of our knowledge, no comprehensive and systematic survey of these efforts exists in the literature.

This work aims precisely at offering an exhaustive overview of the solutions proposed for the application placement problem in the Fog, by providing the reader with \uib{two} complementary perspectives on this topic:
%. Namely:
\begin{itemize}
\item[P1.] an \textit{algorithmic} perspective that reviews state-of-the-art contributions based on the methodologies that they employed to address Fog application placement, along with a study on the available prototypes and experiments, and
\item[P2.] a \textit{modelling} perspective that analyses which (functional and non-functional) constraints and which optimisation metrics have been considered in the literature to determine best candidate application placements.
%\item[P3.] a \textit{future work} perspective that, based on both P1 and P2, identifies and discusses some of the open research challenges that should be studied on this topic.
\end{itemize}
%\noident \uib{All the studies included in this paper are analysed and classified under these two perspectives. Finally, based on both P1 and P2, some of the open research challenges that should be studied on this topic are identified and discussed.}

\medskip\noindent
The rest of this paper is organised as follows. After describing the methodology followed to realise this survey (Section \ref{settingthestage}), a comprehensive analysis of state-of-the-art related to Fog application placement under P1 (Section \ref{sect:algorithms}) and P2 (Section \ref{modelling}) is presented. Finally, based on both P1 and P2, some open problems and future research challenges  are pointed out (Section \ref{openchallenges}).

\section{Setting the Stage}
\label{settingthestage}

This survey includes research articles that deal with Fog application placement with the objective of optimising non-functional requirements of the system. To set the stage, we start by formally defining the considered application placement problem.

\medskip
\textbf{Definition} -- Let $A$ be a multi-component \uib{(or multi-service)} application with a set of requirements $R$ and let $I$ be a distributed Fog infrastructure. Solutions to the \textit{Fog Application Placement Problem} (FAPP) are mappings from each component of $A$ to some computational node in $I$, meeting all requirements set by $R$ and optimising a set of objective metrics $O$ used to evaluate their quality. Solution mappings can be many-to-many, i.e. a component can be placed onto one or more nodes and a node can host more than one component.

%\bigskip

%  \noindent It is worth noting that FAPP can be solved or adjusted also at runtime (i.e., when $A$ is running), in case some requirements in $R$ cannot be met by the current solution mapping. In what follows, focusing on existing approaches to solve FAPP, we will exclude those works that only deal with dispatching or scheduling of (user or IoT) requests. Clearly, they represent a phase that is subsequent to the placement of the application components. Similarly, we will exclude research in the field of task offloading (i.e., outsourcing of the computation of a given function to get a result back) from one node to a better one (e.g., more powerful or reliable, closer), which is covered in detail by other recent surveys (e.g., \cite{aazam2018,xu2018}). 

%\noindent\textbf{Definition} -- Let $A$ be the set of multi-component applications that are deployed in the system with (hardware, software and QoS) requirements $R$. The applications are defined as a set of interacting modules or components $M$. Let $I$ be a distributed (Fog) infrastructure with interconnected nodes $N$ with computational capacities. Solutions to the \textit{Fog Application Placement Problem} (FAPP) are mappings from each component in $M$ to computational nodes in $N$, meeting all requirements set by $R$. Solution mappings can be many-to-many, i.e. a component can be placed onto one or more nodes and a node can host more than one component. Those solutions are addressed to optimise a set $O$ of metrics of the systems, while the considered constraints $C$ are satisfied.

\medskip
  \noindent It is worth noting that FAPP can be solved or adjusted also at runtime (i.e., when $A$ is running), in case that some requirements in $R$ cannot be met by the current solution mapping or whenever $O$ can be further optimised as Fog infrastructure conditions change over time. 

The concept of FAPP and the underlying technologies are currently emerging, and the boundary with other technologies is not always clear. In this survey, we include all the articles that deal with the placement of applications that are generally available for the users and that \uib{can benefit from the adoption of the Fog computing paradigm.}
%have been traditionally requested to Cloud providers. 
%
Focussing on existing approaches to solve FAPP, we will exclude those works that only deal with dispatching or scheduling of (user or IoT) requests, as they represent a phase that is subsequent to the placement of the application components.
Similarly, we will exclude research in the field of task offloading (i.e., outsourcing of the computation of a given function to get a result back) from one node to a better (e.g., more powerful or reliable, closer) one, which is covered in detail by other recent surveys\cite{aazam2018,xu2018}. 

\uib{A problem similar to FAPP has been studied in the context of Software Defined Networking (SDN), and literature on the placement of Virtual Network Functions (VNFs) onto SDN substrates has been already thoroughly analysed in exhaustive recent surveys (e.g.~\cite{8066287,KARAKUS2017200,ASSEFA2019127,SALMAN2018221}). Such a problem is known as \textit{VNF embedding} and includes the joint placement of virtual function services and the routing of traffic flows between them. FAPP significantly differs from VNF embedding as the latter assumes to have the possibility of programming network flows from a (logically) centralised point of control in the infrastructure. As Fog application deployments will likely span various service providers, these might not always support SDN across their infrastructures, hence the previous assumption might be too stringent when considering Fog scenarios in general. Thus, also to avoid overlapping with existing surveys on SDN and VNF placements, in what follows we will exclude results in those areas.
%commit our efforts to analyse the literature focussing on FAPP only.
}
%\uib{Works related to the use of software-defined networks to improve the QoS of Fog applications will be also excluded. This field is also analysed in respective surveys~\cite{8066287,KARAKUS2017200,ASSEFA2019127}} 
%
Finally, research in the field of mobile offloading, shifting the processing and execution from mobile terminals to network capabilities or edges nodes, will be also excluded. 

\medskip
To the best of our knowledge, this is the first survey that covers the state of the art for FAPP. Our search criteria were formed by the following terms: {\sf (Fog computing $\wedge$ (service $\vee$ application) $\wedge$ (placement $\vee$ deployment))}. 
The search was carried out, with the help of Google Scholar, in the following libraries: IEEE Xplore Digital Library, Wiley Online Library, ACM Digital Library and Web of Science. To fully capture the advances in the field, both journal and conference articles were collected during the search phase. Additionally, the references in selected articles were also analysed to find more related work in the FAPP domain. 
At the end of this first step, the 110 articles we collected were carefully screened. After a more accurate and deeper selection process, where references in the field of offloading, dispatching, scheduling and VNF embedding were removed, we selected 38 articles to be further analysed. 

\medskip\noindent
Table~\ref{tab:FAPPsummaryalgorithm} offers a complete bird's-eye view of the set of the analysed papers highlighting their main strengths and weaknesses, and the availability of an open-source prototype implementation. Focussing the definition of FAPP from an optimisation point of view, the common elements in any optimisation process are the following:

\begin{description}
\item{\bf Decision variables} -- They are the items whose values need to be determined during the optimisation process. In the case of the FAPP, they can be the binary decision variables (or, equivalently, the mapping functions) that indicate if a component of $A$ is allocated (or not) to a Cloud or Fog node of $I$. All the articles that we have included in our study, rely on similar decision variables.

\item{\bf Objective function} -- The objective function measures the suitability of a solution (a specific value assigned to the decision variables) within the optimisation process. The optimisation can be addressed to maximise or minimise the value of the objective function. In a more general way, the objective function represents the concerns of the optimisation and defines the metrics that are optimised by fixing the values of the decision variables. In our case, the objective functions measure the metrics $O$ of candidate solutions to FAPP. 

\item{\bf Constraints over the decision variables} --  They define the requirements $R$ that must be satisfied by each specific case of the decision variables. Solutions (values of the decision variables) that do not satisfy the constraints are rejected as possible solutions to FAPP.

%They are presented in Section~\ref{sect:constraints}.

\item{\bf Domain variables or parameters}  -- Commonly, they are the fixed values which are known previously to the optimisation process. We also include in this category other variables, metrics or items related to the optimisation that are not constraints neither optimisation objectives. 

%They are also presented in Section~\ref{sect:constraints} along with the constraints.

%\uib{An overview of the set of the analysed papers with respect to all the factors that were included in our survey of the state of the art is offered in Figure~\ref{fig:algorithmbasedclassification} (algorithms),  Table~\ref{tab:FAPPsummaryalgorithm} (algorithms, solution's strengths and weaknesses, and availability of an open prototype), Table~\ref{tab:FAPPsummaryconstraints} (model constraints), and Table~\ref{tab:FAPPsummarymetrics} (optimisation metrics).} Focussing the definition of FAPP from an optimisation point of view, the common elements in any optimisation process are the following:

\item{\bf Algorithms}  -- They are the algorithms proposed to find the values of the decision variables that optimise the objective function value, while satisfying all set constraints. 
The problem of finding solution mappings between application components in $A$ and Fog/Cloud nodes in $I$, whilst minimising/maximising the values of the objectives functions $O$ and satisfying the constraints in $R$, is NP-hard. Moreover, when several opposite optimisation objectives are considered, it is necessary to determine a trade-off between the objectives because some of them may increase when the other decrease. Indeed, FAPP needs to be solved by evaluating -- at worst -- all possible solutions, i.e., assuming the application is made of $m$ modules and the infrastructure is composed of $n$ nodes, $n^{m}$ different candidate solutions. 
Such complexity can be tamed by using heuristics which permit to find sub-optimal solutions. It is also important to consider that Fog computing is a large-scale and dynamic domain, where real implementations have to deal with huge quantities of Fog nodes and applications.  
\end{description}

 \begin{table}
% \begin{minipage}{\columnwidth}
	\centering
	\caption{\uib{Overview of the analysed articles: algorithms, strengths, weaknessess and available prototypes.}}\label{tab:FAPPsummaryalgorithm}

	%Scale the table to fit on the page: or this line or the next
    \resizebox{0.99\columnwidth}{!}{
   % \scalebox{0.6}{
     \rowcolors{3}{gray!20}{white}
     \begin{tabular}{@{}|c|c|p{0.7\columnwidth}|p{0.7\columnwidth}|c|@{}}\toprule
  	%\begin{tabular}{@{}cccccccccccccccccccc@{}}\toprule
%	\begin{tabular}{c|c|c|c|c|c|c|c|c|c|c|c|c|c|c|c|c|c|c|c|c|c}\toprule
         Ref. & Alg.\footnote{Algorithm: S Search; MP Mathematical Programming; BI Bio-inspired; GT Game Theory; DL Deep Learning; DP Dynamic Programming; CN Complex Networks} & Strengths & Weaknesses & Open  \\
          & &  &  &   Proto. \\

        \midrule
%1 &2 & 3 & 4 & 5 & 6 & 7& 8 & 9 & 10 & 11 &  12 & 13 & 14 & 15 & 16 & 17 & 18 & 19 & 20& 21\\

%% SEARCH ALGORITHMS
\cite{004}\cite{bookchapterbuyya} & S &Heuristic search. Modelling of IoT, Fog and Cloud. Extension to a well-known Cloud simulator. Medium-scale experiments (up to 80 Fog nodes). &Static tree-based infrastructures and DAG applications only. No complexity analysis. & \checkmark\\
\cite{021} & S & First-fit service placement and distributed \textit{a posteriori} optimisation of delays and idle resource usage. & Static tree-based infrastructures and DAG applications only. Lack of details on possible real-world implementation. Small-scale example (up to 30 Fog nodes).& \\
%18
\cite{101} & S &Distributed placement strategy based on local data of each node. Validated on a real application topology. Improved network usage with respect to users demand. & Static tree-based infrastructures and DAG applications only. Increased number of service migrations. Small-scale experiments (up to 25 Fog nodes). &\\
%20
\cite{036} & S & Heuristic search for best placement and migration plan. Uncertainty in user mobility and data rates is considered. Large-scale example (up to 1000 Fog nodes). &Use of simple prediction mechanisms. Cubic-time complexity.& \\

%26
%TODO Antonio & Stefano
\cite{011} & S & Heuristic search. Arbitrary application and infrastructure topologies, modelling of IoT and asymmetric network QoS. Complexity proofs.  & Static infrastructure conditions. Small-scale example (5 Cloud and Fog nodes). Worst-case exp-time. & \checkmark\\
%27
%TODO Antonio & Stefano
\cite{012} & S & Exhaustive search. Dynamic network QoS through Monte Carlo simulations. Possibility to perform \textit{what-if} analyses.  & Small-scale example (5 Cloud and Fog nodes). Exp-time complexity. & \checkmark\\
%28,29
%TODO Antonio & Stefano
\cite{102}\cite{bookchapter} &S& Parallel exhaustive search. Cost model considering operational IoT, Fog and Cloud costs. Possibility to perform \textit{what-if} analyses. Medium-scale examples (up to 80 Fog nodes). &  Exp-time complexity (tamed by multi-thread implementation). & \checkmark \\

\cite{109} &S&Heuristic search. Arbitrary application and infrastructure topologies. High scalability and large-scale experiments (up to $20000$ nodes). Comparison with exhaustive and first-fit.& Static infrastructure conditions. No codebase released. & \\

%29
\cite{032} & S &Exhaustive search. Medium-scale experiments (50 nodes). Hints for a distributed extension to the approach. & Linear application chains and static tree-based infrastructure only. Quadratic-time complexity.& \\
% Search 31
\cite{062} & S &Heuristic search. Online implementation with Kubernetes. Linearithmic-time complexity. Real laboratory testbed. &Only considers generic device capacity. Does not consider infrastructure topology nor network QoS. Small-scale experiments (5 Fog nodes).& \\
%32
\cite{027} & S &Heuristic search. Linearithmic time complexity. Preliminary hints for a benchmark. & Static tree-based infrastructures and DAG applications only. Small-scale example (up to 13 Fog nodes). & \\

%% ILP ALG
 \cite{029} &MP&Detailed framework and architectural proposal for solving FAPP with ILP. & No implementation nor evaluation is available. & \\

%35
\cite{001} &MP&
MINLP and linearisation into MILP. Joint solution of FAPP and task distribution to minimise operational costs. Medium-scale experiments (up to 100 Fog nodes). & %flabby
Data trace of the experiments are not provided. Static infrastructure conditions.
& \\

%36
\cite{034} &MP&
ILP and problem relaxation. Workload variations according to users mobility. Data migration costs. Small-scale example (20 nodes), comparison with GA and classical ILP.& Cubic-time complexity. Static tree-based infrastructures and DAG applications only. & \\

\cite{017} &MP& MINLP and linearisation into MILP. Joint solution of FAPP and task distribution to minimise response times. Medium-scale example (up to $80$ nodes).
& No  complexity analysis provided. Complex formulation. & \\

%37
\cite{095} &MP& MINLP and linearisation into MILP. Comparison with greedy strategy. & No  complexity analysis provided. Static infrastructure conditions. Small-scale example ($\simeq 10$ nodes). Not very detailed experimental settings.& \\

%38
\cite{061} &MP& ILP. Optimising latency for resource allocation. & Only models two application types and generic device capacity. Small-case example (6 Fog nodes).& \\

%39
\cite{002} &MP&
Mixed-cast Flow Problem. Joint solution of FAPP and requests routing to minimise operational costs. Mock data traces used for experiments. 
& Not very detailed experimental settings.  & \\

%&Heuristic search. Modelling of IoT, Fog and Cloud. Extension to a well-known Cloud simulator. Medium-scale experiments (up to 80 Fog nodes). &Static tree-based infrastructures and DAG applications only. No complexity analysis. &
%40
\cite{107} &MP& ILP. QoE-placement, aware of (fuzzy) user expectations. Comparison with other 3 heuristic algorithms. & Static tree-based infrastructures and DAG applications only. No complexity analysis provided. & \\

\cite{024,066} &MP& ILP. Distributed management of Fog colonies (infrastructure portions). Modelling of IoT, Fog and Cloud nodes. & Static tree-based infrastructures. Simplistic linear cost model. Small-scale example. & \\

 \cite{025} &GA& GA. Distributed management of Fog colonies (infrastructure portions). Modelling of IoT, Fog and Cloud nodes. Poly-time complexity. Comparison with first-fit and ILP strategies. & Static tree-based infrastructures. Fitness function based on constraints satisfaction instead of optimisation metrics. Small-scale example (10 Fog nodes). & \\

%47
%\cite{066} &MP&Distributed management via Fog colonies/partitions. Use of a real network topology.& Experiments with only one Fog colony. Cost evaluation neglects the Fog execution costs.& \\
\cite{019} &MP& ILP. Heuristic to solve an equivalent problem. Considering communication energy consumption. Complexity proof. Large-scale experiments (1000 Fog nodes). & Experiments focus only on energy results. & \\
%49
\cite{014} &MP& LP. Decomposition of the primal problem into sub-problems to optimise the power consumption-delay tradeoff. &  Static infrastructure conditions. Small-scale example (8 nodes).& \\

%51
\cite{030} &MP& ILP. Management of Fog colonies (infrastructure portions). Extensive comparison of Cloud-only, Fog-only and Fog-to-Cloud placements. Real smart-city scenario. & Static tree-based infrastructures.  Small-scale experiments (1 Cloud, 1 Fog node). & \\
%54-%55

\cite{105}\cite{106}\cite{038} &MP& ILP. Heuristic QoS-aware extension of Apache Storm distributed framework. Real small testbed (8 nodes). Medium-scale simulation (up to 100 Fog nodes). Complexity proof. &
Worst-case exp-time. Placement policy might deteriorate application availability when many operators are involved.
&

\checkmark\\

 %&MP&Rigorous mathematical formulation of distributed stream processing applications. Integration in Apache Storm,& Run-time not supported in terms of a properly handling of input streams or execution environments with continuously changing properties.& \\

%% OTHER ALGORITHM
%60
\cite{033} &BI& GA. Perspective and future directions in FAPP with a focus on GA and their parallel implementation. & Preliminary work. Simple case study.
 & \\
%61
\cite{076} &BI& GA. Distributed strategy for handling application replicas and guaranteeing reliability against probabilistic infrastructure failures. Comparison with ILP. & Limited scalability. Small-scale example (5 Fog nodes). &   \\

\cite{053} &GT& Stackelberg games. FAPP modelled as sub-problems of pairing subscribers to needed resources, and of deciding resource prices. & No dynamic conditions considered. Small-scale example (20 Fog nodes). & \\

%64
\cite{088} &GT& Stackelberg games. Pricing analysis using Stackelberg game in a three-tiered network model (student-project allocation algorithm).& No dynamic conditions considered. Some details missing on experimental settings. & \\
%66
\cite{008} &DL& Q-learning. Service migration and user mobility are considered. Medium-scale experiments (70 nodes) with real-world driver traces. & Complex formulation. Cubic-time complexity. Parameter tuning not very clear. & \\

%67
\cite{064} &DP&0-1 Knapsack. Load and energy adaptive strategy according to the resource availability. &No algorithmic details. Small-scale example (9 nodes). & \\

%69
\cite{006} &DP& Knapsack problem. Exploratory work in symbiotic search. Comparison with FCFS policy and traditional solvers. & Very small-scale example.& \\

\cite{110} &CN&Network science. Mobility of users is considered. Extension to a well-known Cloud simulator. & Static tree-based infrastructures. Medium-scale example (80 Fog nodes). & \\
\bottomrule
	\end{tabular}
}
%\end{minipage}
\end{table}

\medskip
\noindent Overall, we have organised the description and presentation of the articles from the point of view of the elements of the optimisation process: algorithms (Section~\ref{sect:algorithms}), constraints and parameters (Section~\ref{sect:constraints}), and objective functions (Section~\ref{sect:objective}). 
\uib{Table \ref{tab:FAPPsummaryalgorithm} (and Fig.~\ref{fig:algorithmbasedclassification}), Table~\ref{tab:FAPPsummaryconstraints}, and Table~\ref{tab:FAPPsummarymetrics} give an overview of the characteristics of the analysed articles according to such three elements, respectively:}

\begin{description}
    \item{\it Algorithms view point} --
\uib{The analysis of the selected articles has shown that various algorithmic approaches have been proposed to optimise FAPP \uib{(Fig. \ref{section31Fig})}. We have grouped and analysed the articles in Section~\ref{sect:algorithms} by those optimisation algorithms resulting in two main groups, with the highest number of articles proposing search-based and mathematical programming solutions to FAPP. Other types of algorithms -- i.e., bio-inspired, game theoretical, deep learning, dynamic programming, and complex networks algorithms -- have been only studied in a more limited number of works.}

    \item{\it Constraints view point} --
In the case of the constraints, we analysed the articles and classified the constraints into a two-level taxonomy \uib{(Fig. \ref{taxonomyconstraints})}. \uib{In a first level, we considered constraints related to the main elements of Fog architectures, i.e. \textit{network} constraints, characteristics of the available Fog \textit{nodes}, \textit{energy} constraints, and \textit{application} requirements. By further analysing the articles, we detected that network constraints related to the available Fog network topology (e.g. availability of IoT devices, existing communication links), to the experienced end-to-end \textit{latency} (or communication time) among nodes, to \textit{bandwidth} availability, and to \textit{link reliability}. We also sub-divided node-related constraints relating them to the availability of \textit{hardware} resources (e.g., RAM, CPU, storage) and of particular \textit{software} frameworks (e.g., libraries, OSs). Similarly, application constraints are related with the type and number of requests that are generated in the systems (\textit{workload}), to the organisation and interrelation between application components or services (\textit{dependencies}), and to the possibility for the users to set conditions in the application placement or to choose between a set of alternative placements (\textit{user preferences}). Lastly, energy constraints are only related to the power consumption and a second classification was not considered for them.}

    \item{\it Objective functions view point} --
\uib{From the analysis of the optimisation objectives, we observed that the articles could be again described by a two-level taxonomy (Fig. \ref{taxonomymetrics}). Particularly, optimising \textit{network} usage was a common objective, attempting to reduce communication \textit{delay} (often including processing times) and network \textit{bandwidth} consumption. Similarly, many approaches attempted an optimisation on the amount of \textit{hardware} resources allocated to application deployment and to energy objectives, related to the power consumption that the execution of applications generates in the Fog layer. With respect to system \textit{performance}, the optimisation focus was on overall \textit{QoS-assurance} (e.g., measured as the number of requests executed before a specific deadline or as the likelihood to satisfy certain constraints on QoS requirements), on the \textit{execution time} of the application requests, or on the overhead due to \textit{migrations} of application components or services from different Fog or Cloud nodes. Finally, operational \textit{cost} (due to hardware, software and bandwidth purchase) for keeping an application deployment up and running was often considered among the optimisation objectives.}
\end{description}

\section{Analysis of the State of the Art}
\label{analysis}

\subsection{Algorithms}
 \label{sect:algorithms}

%TODO DOUBLE-CHEK: Design?, reference in text?, caption?

%UPI

\begin{figure}[]
\centering
\includegraphics[width=10cm]{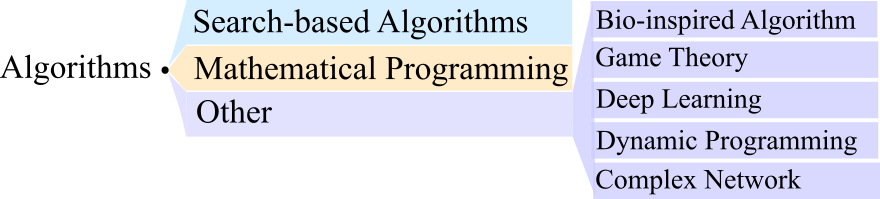}
\caption{\uib{Taxonomy of FAPP algorithms.}}
\label{section31Fig}
\end{figure}

In this section, all reviewed approaches are analysed according to the algorithms or methodologies that they use for solving FAPP. 
As aforementioned, we identified three main classes of algorithms which have been exploited in the literature \uib{(Figure~\ref{section31Fig})}. Namely:

\begin{itemize}
\item \textit{search-based} algorithms, such as (heuristic) backtracking search, or first- and best-fit application placement, %or genetic algorithms,
\item \textit{mathematical programming} algorithms, such as integer or mixed integer linear programming,
\item \textit{other} algorithms like game theoretical or deep learning.
\end{itemize}

\begin{figure}[]
\centering

\begin{tikzpicture}[show/.style={circle,draw}]
\node[show, align=center]    (search)    at    (2,4)    
    [label=above:{Search-based Algorithms}]    
    {\cite{004} \cite{bookchapterbuyya} \cite{021}\\ \cite{101} \cite{036} \cite{011} \cite{012}\\ \cite{102} \cite{bookchapter} \cite{109}\\ \cite{032} \cite{062} \cite{027}};
\node[show, align=center]    (mp)   at     (0,1)    
     [label=below:{Mathematical Programming}]    
    {\cite{029} \cite{001} \cite{034} \cite{017}\\ \cite{095} \cite{061} \cite{002} \cite{107} \cite{024}\\ \cite{066} \cite{019} \cite{014}\\ \cite{030} \cite{105} \cite{106} \cite{038} };
\node[show,minimum size=4cm]    (other)   at     (4,1)    
     [label=below:{Other Algorithms}]    
    {};
\node[show]    (bio)   at     (4,2) 
     [label={[label distance=-3pt]above:{\tiny Bio-inspired}}]   
    {\cite{025} \cite{033} \cite{076}};
    
\node[show]    (game)   at     (3,1) 
     [label={[label distance=-3pt]above:{\tiny Game Theory}}]    
    {\cite{053} \cite{088}};
    
\node[show]    (deep)   at     (4,1) 
     [label={[label distance=-3pt]below:{\tiny Deep Learning}}]    
    {\cite{008}};
    
\node[show]    (dynamic)   at     (5,1) 
     [label={[label distance=-3pt]above:{\tiny Dynamic Prog.}}]    
    {\cite{064} \cite{006}};
    
\node[show, align=center]    (cn)   at     (4,0) 
     [label={[label distance=-3pt]below:{\tiny Complex Networks}}]    
    {\cite{110}};

\end{tikzpicture}

\caption{\uib{Overview of the analysed articles by employed algorithm.}}
\label{fig:algorithmbasedclassification}
\end{figure}
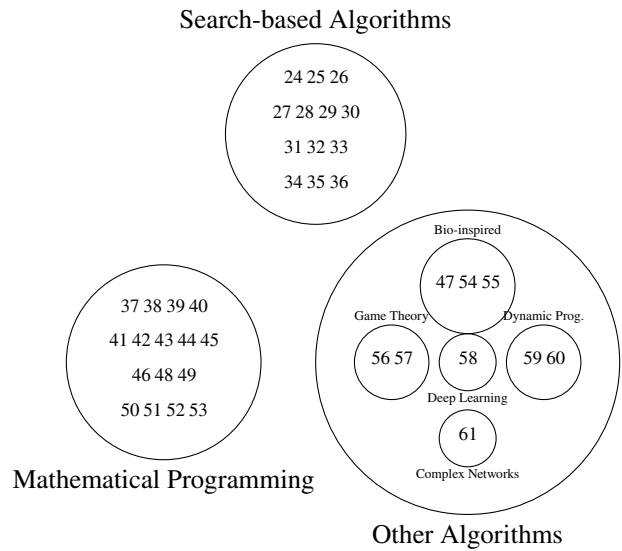

\noindent 
\uib{Figure~\ref{fig:algorithmbasedclassification} graphically depicts the classification of the surveyed algorithms in  Table~\ref{tab:FAPPsummaryalgorithm} from such an algorithmic perspective.}
The rest of this section is organised according to this classification with the goal of providing the reader with a complete overview of the state of the art related to frameworks used for solving FAPP (Sections \ref{search_algs} to \ref{other_algs}). For each reviewed approach, we indicate the availability of open source research prototypes and the size of the experiment used for assessing or testing it. \uib{Strengths and weaknesses of each work are also summarised in Table~\ref{tab:FAPPsummaryalgorithm}.}

\subsubsection{Search-based Algorithms}\label{search_algs}

Being FAPP an NP-hard problem, the first solutions which have been exploited to solve it are traditional search algorithms along with greedy or heuristic behaviour \cite{aima}. 

\medskip

%IFogSim
Among the first proposals investigating this direction, Gupta et al.\cite{004} and Mahmud et al.\cite{bookchapterbuyya} proposed a Fog-to-Cloud search algorithm as a first way to determine an eligible (composite) application placement of a given (Directed-Acyclic) IoT application to a (tree-structured) Fog infrastructure. The search algorithm proceeds Edge-ward, i.e. it attempts the placement of components \textit{Fog-to-Cloud} by considering hardware capacity only, and allows merging homologous components from different application deployments. An open-source tool -- iFogSim -- implementing such strategy has been released and used to compare it with \textit{Cloud-only} application placement over two quite large use cases from VR gaming (3 application components scaled up to 66, 1 Cloud, up to 80 Fog nodes) and video-surveillance (4 application components, scaled up to 19, 1 Cloud, up to 17 Fog nodes). \uib{Building on top of iFogSim, Mahmud et al.\cite{021} refined the Edge-ward algorithm with an \textit{a posteriori} management policy to guarantee the specified application service delivery deadlines and to optimise Fog resource exploitation. They illustrated their proposal at work on two small-scale motivating examples (up to $30$ Fog nodes).}

\medskip
%UIB
Recently, exploiting iFogSim, Guerrero et al.\cite{101} proposed a distributed search strategy to find the best service placement in the Fog, which minimises the distance between the clients and the most requested services. Each Fog node takes local decision to optimise the application placement, based on request rates and freely available resources. The \textit{SocksShop} \cite{weaveworks2016shocksshop} application (9 components) is used to evaluate the proposed decentralised solution against the Edge-ward policy of iFogSim --varying the replication factor in the range $1-5$  and the number of Fog nodes in the range $1-25$. The results showed a substantial improvement in network usage and service latency for the most frequently requested services. Also, Ottenwalder et al.\cite{036} proposed a distributed solution -- MigCEP -- both to FAPP and to runtime migration. Such proposal reduces the network usage in Fog environments and ensures end-to-end latency constraints by searching for the best migration, based on probabilistic mobility of application users. The OMNeT++ \cite{omnet} simulator was used to show how MigCEP improves live migration against a static and a greedy strategy over a dynamic large-scale infrastructure (i.e., 1000 emulated connected cars). No codebase was released for either the works of Guerrero et al.\cite{101} or Ottenwalder et al.\cite{036}.

\medskip
%UPI
In this context, Brogi and Forti\cite{011} proposed both an exhaustive and a greedy backtracking algorithm to solve FAPP based on the (hardware, software, IoT and QoS) requirements of multi-component applications. The greedy heuristic attempts the placement of components sorted in ascending order on the number compatible nodes (i.e., \textit{fail-first}), considering candidate nodes one by one sorted in decreasing order on the available resources (i.e., \textit{fail-last}). The devised approach works on arbitrary application and infrastructure graph topologies. Later on, Brogi et al. extended their initial model and algorithms probabilistically so to estimate QoS-assurance, resource consumption in the Fog layer \cite{012} and monthly deployment cost~\cite{102} of eligible placements. 
QoS-assurance is estimated by means of Monte Carlo simulations so to consider variations in the QoS of end-to-end communication links and to predict how likely each application placement is to comply with the desired network QoS \cite{012}. An open-source prototype -- FogTorch$\Pi$ -- implementing the whole methodology has been released and used on different simple use cases (i.e., 3 application components, 2 Clouds, 3 Fog nodes) from smart agriculture and smart building scenarios. Despite exploiting worst-case exponential-time algorithms, the prototype was shown to scale \cite{bookchapter} also on the larger VR game example proposed by Gupta et al.\cite{004}. FogTorch$\Pi$ was also modularly extended by other researchers to simulate mobile task offloading in Edge computing \cite{demaio}.

\medskip
Significantly inspired by Brogi and Forti\cite{011}, Xia et al.\cite{109} proposed a backtracking solution to FAPP to minimise the average response time of deployed IoT applications. Two new heuristics were devised. The first one sorts the nodes considered for deploying each component in ascending order with respect to the (average) latency between each node and the IoT devices required by the component. The second one considers a component that caused backtracking as the first one to be mapped in the next search step. A motivating IoT application \cite{doorbell} (i.e., 6 application components replicated up to $\simeq$ 800 times) was used to assess the algorithms on very large random infrastructures (i.e., 1 Cloud, up to $\simeq$ 20000 Fog nodes). The exhaustive search handled at most 150 nodes, whilst the first-fit and the heuristic strategy scaled up to 20000 nodes, the latter showing a 40\% improvement on the response time with respect to first-fit. Prototype implementations were not released.

\medskip
%similar to UPI
Limiting their work to linear application graphs and tree-like infrastructure topologies, Wang et al.\cite{032} described an algorithm for optimal online placement of application components, with respect to load balancing. The algorithm searches for cycle-free solutions to FAPP, and shows quadratic complexity in the number of considered computational nodes. %Indeed, the solution relies on auxiliary data structures containing the costs of all pairwise mappings of a component to a given node, which are then searched for the best placement. 
An approximate extension handling tree-like application is also proposed, which considers placing each (linear) branch of the application separately and shows increased time complexity. The approach is simulated on 100 applications to be placed featuring 3 to 10 components each and infrastructures with 2 to 50 nodes. Finally, the achieved performance is compared against the Vineyard algorithm \cite{vineyard} and a greedy search minimising resource usage.

\medskip
Hong et al.\cite{062} proposed a (linearithmic) heuristic algorithm that attempts deployments by prioritising the placement of smaller applications to devices with less free resources. A face detection application made from 3 components is used to evaluate the algorithm on a small real testbed (1 server acting as Cloud, 5 Fog nodes). Taneja and Davy\cite{027} proposed a similar search algorithm that assigns application components to the node with the lowest capacity that can satisfy application requirements, also featuring linearithmic time complexity in the number of considered nodes. Binary search on the candidate nodes is exploited as a heuristic to find the best placement for each component, by attempting deployment to Fog nodes first (i.e., Fog-to-Cloud). The medium-scale experiments (i.e., 6 application components, 1 Cloud, up to 13 Fog nodes) were carried in iFogSim, but the code to run them has not been made publicly available.

\medskip

\subsubsection{Mathematical Programming}\label{lp_algs}

Mathematical programming \cite{mathematicalprogramming} is often exploited to solve optimisation problems by systematically exploring the domain of an objective function with the goal of maximising (or minimising) its value, i.e., identifying a best candidate solution. Many of the reviewed approaches tackled FAPP with such a mathematical framework, by relying on Integer Linear Programming (ILP), Mixed-Integer Linear Programming (MILP) or Mixed-Integer Non-Linear Programming (MINLP).

\medskip
Velasquez et at.\cite{029} proposed a framework and an architecture for application placement in Fog computing. An ILP implementation is suggested but it is not realised nor evaluated by the authors. On the other hand, Arkian et al.\cite{001}, in addition to proposing a Fog architectural framework, formulated FAPP as a MINLP problem, where application components (i.e., VMs) are to be deployed to Fog nodes so to satisfy end-to-end delay constraints. The problem is then solved (along with the problem of task dispatching) by linearisation into a MILP and the solution is evaluated on data traces from IoT service demands (from 10-100 thousand devices) in the province of Teheran, Iran. The results of the experiment showed that the Fog promises to improve latency, energy consumption and costs for routing and storage. 
Similarly, Yang et al.\cite{034} tackled both FAPP and its cost-aware extension with the problem of balancing request dispatching. The proposed methodology attempts optimising the trade-off between access latency, resource usage, and (data) migrations (costs). It accounts for constraints on the available resources as well as for workload variations depending on users' service accessing patterns. A novel greedy heuristic (solving a relaxed LP problem and approximating a solution for the full-fledged ones) is shown to outperform other benchmark algorithms (i.e., classic ILP and GA) both in terms of obtained results and algorithm execution time over an example with 20 Fog nodes and 30 services to be placed.

\medskip
Alike to these proposals, Zeng et al.\cite{017, 095} solved FAPP along with task scheduling, converting a MINLP into a MILP, solved with the commercial tool Gurobi \cite{gurobi}. A simulation and comparison is provided against server-greedy (i.e., Cloud-ward) and client-greedy (i.e., Edge-ward) placement strategies, showing good improvements on response time when using the proposed approach in a small (i.e., 15 to 25 application images, 11 Fog/Cloud servers\cite{095}) and medium (i.e., 45 to 80 Fog nodes\cite{017}) sized use case.
Still relying on Gurobi (together with PuLP \cite{mitchell2011pulp}), Souza et al.\cite{061} solved FAPP as an ILP problem, aimed at optimising latency/delay needed for resource allocation. Resources are modelled uniformly as available slots at different nodes. Out of the services to be deployed, few of them were considered to require large amounts of resources (i.e., \textit{elephants}, 10\%) and many more required instead few resources (i.e., \textit{mice}, 90\%). In the small-scale experimental settings (90 applications, 1 Cloud, 6 Fog nodes) both sequential and parallel resource allocation were considered and the benefit of Fog computing was shown in terms of reduced delays in service access. 

\medskip
Alternatively, Barcelo et al.\cite{002} combined FAPP with the routing of requests across an IoT-Cloud (i.e., Fog) infrastructure. They modelled FAPP as a minimum (energy) cost mixed-cast flow problem, considering unicast downstream and multicast upstream, typical of IoT. A solution to FAPP is then provided by means of known poly-time algorithms, which are simulated via the LP solver Xpress-MP \cite{xpressmp} on mock data traces from three use cases (i.e., smart city, smart building, smart mobility). The experiments simulated tens of devices and showed some performance improvements  (as per latency, energy, reliability) with respect to traditional IoT deployments that do not rely on Fog nodes. Mahmud et al.\cite{107} proposed instead a QoE extension of iFogSim -- based on an ILP modelling of users expectation -- which exploited fuzzy logic and achieved improvements in network conditions and service quality.

\medskip
Skarlat et al. designed a hierarchical modelling of Fog infrastructures, consisting of a centralised management system to control Fog nodes organised per \textit{colonies}\cite{024, 025, 066}. Particularly, Skarlat et al.\cite{024} adopted an ILP formulation of the problem of allocating computation to Fog nodes in order to optimise (user-defined) time deadlines on application execution, considering IoT devices needed to properly run the application. A simple linear model for Cloud costs is also taken into account. The proposed approach is compared via simulation to first-fit and Cloud-only deployment strategies, showing good margins for improvement, on a small-scale use case (i.e., up to 80 services, 1 Cloud, 11 Fog nodes). 

\medskip
Some of the methodologies proposed in the literature combine ILP with other optimisation techniques. Huang et al.\cite{019}, for instance, modelled the problem of mapping IoT services to Edge/Fog devices as a quadratic programming problem, afterwards simplified into an ILP and into a Maximum Weighted Independent Set (MWIS) problem. The services are described as a co-location graph, and heuristics are used to find a solution that minimises energy consumption. A (promising) evaluation is performed over large-scale simulation settings (i.e., 50 services, 100 to 1000 Fog nodes). 

 \medskip
Similarly, Deng et al.\cite{014} followed a hybrid approach to model FAPP and to determine the best trade-off between power consumption and network delays (exploiting $M/M/n$ Markov models to describe network capabilities). After decomposing FAPP into three sub-problems (power-delay vs Fog communication, power-delay vs Cloud communication, and minimising WAN delay) balanced workload solutions are looked for exploiting different optimisation methods (i.e., convex optimisation, generalised Benders' decomposition and Hungarian method). A quite large simulation setup in MATLAB \cite{matlab} was used to evaluate the proposed approach (i.e, from 30000 to 60000 nodes).

\medskip
Unfortunately, none of the approaches previously discussed in this section released the code to run the experiments. Conversely, based on the hierarchical model of Skarlat et al. \cite{025}, Venticinque and Amato \cite{030} proposed a software platform to support optimal application placement in the Fog, within the framework of the CoSSMic European Project \cite{cossmic}. Envisioning resource, bandwidth and response time constraints, their approach permits to choose among a Cloud-only, a Fog-only or a Cloud-to-Fog deployment policy, which were evaluated in a small testbed (i.e., 1 Cloud, 1 Fog node) where a composite application (8 components) from Smart Energy scenarios \cite{cosmic_app} was deployed and run over emulated IoT data traces.

\medskip
Finally, Cardellini et al.\cite{105, 106, 038} discussed and released \uib{Distributed Storm (formerly S-ODP)}, an open-source extension of Apache Storm that solves FAPP by means of the CPLEX \cite{ibmcplex} optimiser with the goal of minimising end-to-end application latency and availability of stream-based applications. Extensive experiments (i.e., up to 50 application components and up to 100 Fog nodes) showed the scalability of their ILP approach (with respect to a traffic-aware extension of Storm \cite{xu_cardellini}), which can be easily extended to include bandwidth constraints and a network-related objective function considering network usage, traffic and elastic energy computation.

%\subsubsection{Genetic Algorithms}\label{ga_algs}

\subsubsection{Other Algorithms}\label{other_algs}
\begin{description}

\item[Bio-inspired Algorithms] Genetic algorithms (GAs) implement meta-heuristics to solve optimisation and search problems based on bio-inspired operators such as mutation, crossover and selection \cite{geneticalgorithms}. Naturally, some of the reviewed works exploited such bio-inspired search algorithms to explore the solution space of FAPP, and to solve it. Wen et al.\cite{033} surveyed Fog orchestration-related issues and offered a first description of the applicability of GAs and parallel GAs to FAPP.

Retaking the ILP model of Skarlat et al.\cite{024} based on Fog colonies and hierarchical control nodes, Skarlat et al.\cite{025} also proposed a GA solution implemented in iFogSim and compared to a greedy (first-fit) heuristic and to an exact optimisation obtained with CPLEX, over a small example (i.e., 5 application components, 10 Fog nodes). Whilst the first-fit strategy does not manage to guarantee user-defined application deadlines, both the exact solution and the GA do. On average, the GA solutions are 40\% more costly than the optimal ones, despite guaranteeing lower deployment delays. 

Mennes et al.\cite{076} required the user to provide a minimum reliability measure and a maximum number of replicas for each (multi-component) application to be deployed. They employed a distributed GA, by using Biased Random-Key arrays to represent solutions (instead of binary arrays). The placement ratio (placed/required) is used to evaluate the proposed algorithm, which showed near-optimal results against a small example (i.e., 10 applications, 5 Fog nodes) and faster execution time with respect to ILP solvers.
Regrettably, open-source implementations are not provided for any of the works exploiting GA.

\medskip
\item[Game Theory] 

Game theoretical models \cite{gametheory} -- which describe well multi-agent systems where each agent aims at maximising its profit whilst minimising its loss -- were fruitfully applied to solve FAPP in \cite{053} and \cite{088}.

Zhang et al.\cite{053} modelled FAPP as Stackelberg games between data service subscribers and data service providers, the latter owning Fog nodes. A first game is used to determine the number of computing resource blocks that users should purchase (based on latency requirements, block prices and utility). A second game is used to help providers set their prices so to maximise revenues. A matching game is used to map providers to Fog nodes based on their preferences. And, finally, matching between Fog nodes and subscribers is refined in order to be stable. The proposal was tested in a simulated MATLAB environment considering 120 service subscribers, 4 data providers, and 20 Fog nodes.

Similarly, Zhang et al.\cite{088} describe a game among service providers, Fog nodes and service subscribers. The mapping between the Fog resources and service subscribers is determined to solve a student-to-project allocation problem. During the game, subscribers consider revenues, data transmission costs, providers' costs and latency to evaluate possible assignments. On the other hand, providers consider revenue from subscribers minus the cost of service delay.

\medskip
\item[Deep Learning] Reinforcement learning trains software agents (with reward mechanisms) so that they learn policies determining how to react properly under different conditions  \cite{deeplearning}.  To the best of our knowledge, only Tang et al.\cite{008} exploited recent reinforcement learning techniques to solve FAPP. After defining a multi-dimensional Markov Decision Process to minimise communication delay, power consumption and migration costs, a (deep) Q-learning algorithm is proposed to support migration of application components hosted in containers or VMs. The proposal took into account user mobility and was evaluated over a medium-sized infrastructure (i.e., $\simeq$ 70 nodes) using real data about users mobility taken from San Francisco taxi traces.

\medskip
\item[Dynamic Programming]

Differently from the others, Souza et al.\cite{064} modelled FAPP as a 0-1 multidimensional knapsack problem \cite{cormen}  with the objective of minimising a given objective function. Limited simulation results on a medium-sized example (40 application components, 6 Fog nodes, 3 Clouds) are provided by the authors. However, no details are given on how the solution is computed, and the code to run the experiments is not available.

Also Rahbari et al.\cite{006} modelled FAPP as a knapsack problem, by considering the allocation of application modules to running VMs in a Fog infrastructure. iFogSim was used to simulate the proposed symbiotic organisms search algorithm, showing some improvements in energy consumption and network usage with respect to a First Come First Served allocation policy and traditional knapsack solvers.

\medskip
\item[Complex networks] 
To the best of our knowledge, only Filiposka et al.\cite{110} relied on network science theory to model and study FAPP, employing on a community detection method to partition the available Fog nodes into a hierarchical dendogram\cite{barabasi}. The dendrogram was then used to analyse different community partitions so to find the most suitable set of nodes to support the VMs that encapsulate the applications. The proposal was validated with CloudSim \cite{cloudsim} over a medium-scale use case (i.e., 80 Fog nodes, and from 150 to 250 application components). The proposed community-based extension to CloudSim is, however, not available.

\end{description}

\subsection{Modelling}
\label{modelling}

\uib{In this section, all reviewed approaches are analysed according to a modelling perspective that accounts for both the considered problem constraints (Section \ref{sect:constraints}) and optimisation metrics (Section \ref{sect:objective}). }

\subsubsection{Considered Constraints}
\label{sect:constraints}

\begin{table}
	\centering
	\caption{\uib{Overview of the analysed articles by considered constraints. }}\label{tab:FAPPsummaryconstraints}

	%Scale the table to fit on the page: or this line or the next
    \resizebox{0.5\columnwidth}{!}{
   % \scalebox{0.6}{
     \rowcolors{6}{white}{gray!20}
     \begin{tabular}{@{}|c|cccc|cc|c|ccc|@{}}\toprule
  	%\begin{tabular}{@{}cccccccccccccccccccc@{}}\toprule
%	\begin{tabular}{c|c|c|c|c|c|c|c|c|c|c|c|c|c|c|c|c|c|c|c|c|c}\toprule
    
	       & \multicolumn{10}{c|}{Considered Constraints}   \\
    	\cmidrule{2-11} 
		
%		&& \multicolumn{6}{c|}{Infrastructure} &  Energy &  \multicolumn{3}{|c}{Application}&& \multicolumn{8}{c}{}  \\
%		\cmidrule{3-12} \cmidrule{14-21} 

	      & \multicolumn{4}{c|}{Network}&\multicolumn{2}{c|}{Nodes}&\multicolumn{1}{c|}{Energy}&
	    \multicolumn{3}{c|}{Application} \\
	\cmidrule{2-11}

		Ref. & 
		%Constraints
		\multicolumn{1}{c}{\rotatebox[origin=c]{90}{Latency}} &
		\multicolumn{1}{c}{\rotatebox[origin=c]{90}{Bandwidth}} &
		\multicolumn{1}{c}{\rotatebox[origin=p]{90}{Link Reliability}} &
		\multicolumn{1}{c|}{\rotatebox[origin=c]{90}{Topology}} &
		 
		\multicolumn{1}{c}{\rotatebox[origin=c]{90}{Hardware}}&
		\multicolumn{1}{c|}{\rotatebox[origin=c]{90}{Software}}&
		  
		\multicolumn{1}{c|}{\rotatebox[origin=c]{90}{Energy}}&
		  		
		\multicolumn{1}{c}{\rotatebox[origin=c]{90}{Workload}}&
		\multicolumn{1}{c}{\rotatebox[origin=c]{90}{Dependencies}}&
		
		\multicolumn{1}{c|}{\rotatebox[origin=c]{90}{User preferences}}

		%Optimization Objectives

        \\\midrule
%1 &2 & 3 & 4 & 5 & 6 & 7& 8 & 9 & 10 & 11 &  12 & 13 & 14 & 15 & 16 & 17 & 18 & 19 & 20& 21\\

%% SEARCH ALGORITHMS
\cite{004} \cite{bookchapterbuyya}&\checkmark&\checkmark&&\checkmark&\checkmark&&&&&\\
\cite{021}&\checkmark&&&&\checkmark&&&&\checkmark&\\
%18
\cite{101}&\checkmark&&&\checkmark&\checkmark&&&\checkmark&\checkmark&\\
%20
\cite{036}&\checkmark&\checkmark&&\checkmark&&&&\checkmark&\checkmark&\\
%22
\cite{011}&\checkmark&\checkmark&&&\checkmark&\checkmark&&&&\checkmark\\
%23
\cite{012}&\checkmark&\checkmark&&&\checkmark&\checkmark&&&&\checkmark\\
\cite{102} \cite{bookchapter}&\checkmark&\checkmark&&&\checkmark&\checkmark&&&&\checkmark\\

\cite{109}&&&&&\checkmark&\checkmark&&&&\\

%29
\cite{032}&&\checkmark&\checkmark&&\checkmark&&&&\checkmark&\\

%35 --dck
\cite{062}&&&&&\checkmark&&&&&\\

%32
%\multirow{-11}{*}{\rotatebox[origin=c]{90}{\cellcolor{gray!10} Search }} &\cite{027}&&\checkmark&&&\checkmark&&&&&\\

%% ILP ALG
\cite{029}&&&&&&&&&&\\
%35
\cite{001}&&\checkmark&&&\checkmark&&&\checkmark&\checkmark&\\
%36
\cite{034}&&&&\checkmark&\checkmark&&&\checkmark&\checkmark&\\
\cite{017}&&&&&\checkmark&&&&&\\

%38
\cite{095}&&&&&\checkmark&&&&\checkmark&\\
\cite{061}&\checkmark&&&&\checkmark&&&&&\\
%42
\cite{002}&\checkmark&\checkmark&\checkmark&\checkmark&\checkmark&&\checkmark&&&\\

\cite{107}&&&&&\checkmark&&&\checkmark&&\\
\cite{024}&&&&\checkmark&\checkmark&&&&&\checkmark\\
\cite{025}&&&&\checkmark&\checkmark&&&&\checkmark&\\\

%47
\cite{066}&&&&\checkmark&&&&&&\\
\cite{019}&&&&\checkmark&&&&&&\\
%49
\cite{014}&\checkmark&&&&&&&&&\\

%51
\cite{030}&&\checkmark&&\checkmark&\checkmark&&&&&\\
%54-%55
\cite{105}\cite{106}&&&&&\checkmark&&&&&\\
%\multirow{-16}{*}{\rotatebox[origin=c]{90}{Mathematical Programming }} &\cite{038}&\checkmark&\checkmark&\checkmark&\checkmark&\checkmark&&&&&\checkmark\\

%% OTHER ALGORITHM
%60
\cite{033}&&&&&&&&&&\\
%61
\cite{076}&&\checkmark&\checkmark&&\checkmark&&&&&\\
\cite{053}&\checkmark&&&&&&&&&\\

%64
\cite{088}&\checkmark&&&&&&&&&\\
%66
\cite{008}&\checkmark&\checkmark&&\checkmark&\checkmark&&&&&\\

%67
\cite{064}&\checkmark&&&&\checkmark&&\checkmark&&&\\

%69
\cite{006}&&&&&\checkmark&&&&\checkmark&\\

%\multirow{-10}{*}{\rotatebox[origin=c]{90}{\cellcolor{gray!10} Other Algorithms}} & \cite{110} & &&&\checkmark&\checkmark&&&&&\\
\bottomrule

	\end{tabular}%

}
\end{table}

%TODO DOUBLE-CHEK: Design?, reference in text?, caption?
\begin{figure}[bt]
\centering
\includegraphics[width=10cm]{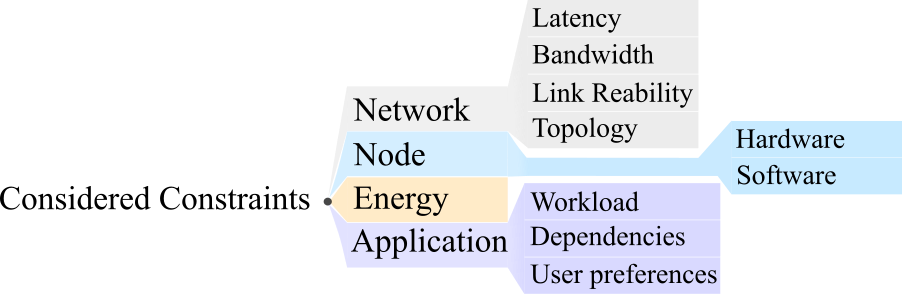}
\caption{\uib{Taxonomy of FAPP constraints.}}
\label{taxonomyconstraints}
\end{figure}

\uib{ 
During the analysis of the modelling of FAPP, we first studied the features of the problem that were taken as constraints to solve it in the surveyed literature. As a result, we obtained the taxonomy of Figure~\ref{taxonomyconstraints} which permits to classify the articles as shown in Table~\ref{tab:FAPPsummaryconstraints}. As aforementioned, the two-level taxonomy of FAPP constraints considers four main types of characteristics, i.e. network features, Fog and Cloud nodes resources, energy consumption and application requirements.}

We defined four features for the second level of the network constraints. Latency is the network feature related with the transmission time of the requests and responses through the network links. This has an important influence on Fog environments since one of the objectives of these domains is to reduce the response time of Cloud-based applications. Consequently, latency is commonly included in the optimisation works targeting FAPP. 
Similarly, the bandwidth -- i.e. quantity of data that the network link is able to transmit per unit of time -- is also important to achieve this objective of reducing the user-perceived response time. In the field of communication networks, the availability and the influence of failures in the transmission is also important, and we additionally included the link reliability in our analysis. 
\uib{Finally, topology information was also included as a criterion so to highlight those articles that considered the distribution and organisation of Fog devices, the influence of the region of the network where the applications are deployed (or users are connected), as well as the presence of IoT devices (i.e., sensors or actuators) that might influence the deployment of application services.}

\medskip
The constraints related to Fog nodes are categorised into hardware and software ones. In the first case, the constraints refer to different types of resources (e.g., processors, memory, storage) or to a generic resource capacity that can represent any considered resource type. In the second case, the constraints are related to software dependencies that should be available at the node hosting a component (e.g., OS, libraries/frameworks, language support).

Energy constraints, commonly related to the power consumption of the hardware elements, are important in Fog domain for two main reasons. The first one is a constraint inherited from Cloud domains. Both Cloud and Fog need a high level of power consumption to give service to the users. Small improvements in power consumption can result on important energy savings. Additionally, in the case of Fog infrastructures, mobile and battery-powered devices are also involved, making energy considerations even more important in the optimisation of FAPP.

Concerning application constraints, we classified the articles by analysing the number of user request (\textit{workload}), the interrelation between the modules of the applications (\textit{dependencies}), and the possibility to express user-defined conditions or preferences related to the application placement (\textit{user preferences}).

\paragraph*{Network Constraints}

The first set of constraints and decision variables are related to the network. The most common metric in this set is the network latency, but others such as the bandwidth and the topology are also quite usual. On the contrary, link reliability is seldom considered in the analysed research works.
Various works solely considered latency among the metrics related with the network\cite{021,061,014,053,088,064}, often but not always along with bandwidth\cite{004,036,011,012,102,bookchapter,002,008}. Bandwidth was also considered in \cite{027,001}, along with hardware constraints. On the contrary, the studies that included link reliability also included other network constraints, such as bandwidth\cite{032,076}, latency and topology\cite{038}, or latency, bandwidth and topology\cite{002}.

\medskip
Topology is the most diverse metric from the ones of the networks as many different proposals exist and have been surveyed in our analyses. Several works deal with the application placement by considering that exist statically defined Fog colonies\cite{024,025,066}, or sets of devices that are managed by a controller node. Thus, a twofold placement based on those colonies was proposed, with a previous mapping of applications in colonies and a second phase of mapping applications to the devices inside a colony. Venticinque and Amato\cite{030} also considered Fog colonies, and they additionally included the network bandwidth constraint.

\medskip
Alternatively, Yang et al.\cite{034} considered that the devices with the capacity to place services are organised in cloudlets, and those cloudlets are assigned to cover a region of the network. A cloudlet is defined as an abstraction tier between the user and the Cloud that provides processing capabilities. It can be a static infrastructure connected to the wireless access network, or augmented routers and switches in the wireless access network. Consequently, the use of a cloudlet would depend on the gateways where the users are connected to and in the topology of the network (reflected in the coverage definition). In this work, the authors also considered the latency of the network as a constraint.

\medskip
Ottenwalder et al.~\cite{036} defined a federation of hierarchy brokers implemented with a combination of Cloud data centres and Fog devices. They also studied the mobility pattern of the users and how they are connected to different nodes in the network topology. The study of the mobility of the users across different parts of the topology of the network is also included in the studies of Tang et al.\cite{008} and Filiposka et al.\cite{110}. Additionally, the latter study used the topological structure of the network as the method to determine the mapping among applications and nodes.
We have also considered that the number of node elements between the users and the applications or between nodes allocating modules of the same application is also a feature related to the topology of the network. Under these conditions, several works\cite{004,101,019} took into account the hop distances of the Fog nodes as input variables of the optimisation process. 

\medskip
Finally, there is a small set of papers that also considered the interactions between applications and IoT devices\cite{004,012,102}, i.e., if the application modules need to be executed in devices with specific hardware requirements or with sensor or actuator elements. We have classified those works into the topology feature since the application placement depends on the characteristics of the network components (nodes).

%Authors in \cite{019} considered the hop distance between devices, and consequently the topology of the network, as an input variable for the optimization of the consumption of the energy of the communication between application services.

%Finally, there is a set of works that consider the hop distance between nodes as a constraint, and this could be interpreted as a topological metric. This is the case with the works in \cite{004} and \cite{101}. 

\paragraph*{Node Constraints}

%In the first case, constraints such as the resource capacity, or the type of hardware are considered. In the second case, the constraints are related to the operating system or tools that would be deployed in the devices. 

The analysis of Table~\ref{tab:FAPPsummaryconstraints}, clearly identified that most of the articles considered constraints about the hardware of the devices. On the contrary, software constraints are only included in a very small number of the surveyed articles on FAPP.

The most common feature related to the hardware was the resource capacity of the Fog devices as a constraint element for the number of services that can be placed into them. The resource capacity is usually modelled as a vector (or set) of elements, one for each of the hardware elements considered in the Fog devices or the network. Examples of those vectors for the case of the Fog devices are: considering CPU\cite{004,008}; considering CPU and storage\cite{001}; CPU, RAM and storage\cite{012,102,017,025}; considering only storage\cite{034,095,024}. Examples of resource capacity vectors both for Fog nodes and network are: considering CPU, RAM and network bandwidth\cite{027,076}; considering CPU, RAM, storage and network bandwidth\cite{109,030}; considering processing and transmission capacities\cite{002}.

\medskip
Other papers defined a general resource model and they did not focus on the specific resource components\cite{021,011,110}. Mahmud et al.\cite{107} also considered a general resource value, but they additionally defined the service demand and device capacity in terms of the expected and offered processing times. On the contrary, this model was sometimes simplified to a scalar value which represents a general capacity unit\cite{101,062,105,106,038}, or with general resources slots\cite{061,064}. Finally, some other papers defined the hardware resources of the Fog computing nodes, but they did not include this constraint in the optimisation process to simplify it\cite{032}.
% * <isaac.lera@uib.es> 2018-07-16T13:48:16.604Z:
% 
% > computing
% Fog computing nodes or Fog nodes?
% 
% ^ <isaac.lera@uib.es> 2018-07-23T10:56:16.117Z.

In other papers, the hardware is considered as variables of a fitness function. The fitness function is used to measure the suitability of a device to place the application modules. Rahbari et al.\cite{006} included the CPU and network usages in the fitness function. For example, Skarlat et al.\cite{024} defined the type of service that determines the additional hardware that the device needs to include to be allocated in.

%Finally, some works additionally consider other hardware constraints as the type of hardware, sensors, or similar that the device is equipped with,~\cite{016}. 

\medskip
As we previously mentioned, constraints related to the software features of the Fog computing nodes are just considered in a small number of studies. For instance, Brogi and Forti\cite{011} and Brogi et al.\cite{012,102,bookchapter} also characterised Fog devices (and application requirements) with software capabilities (e.g., operating system, platforms, frameworks). %Authors in \cite{109} also considered software constraints such as the operating system.

\paragraph*{Energy Constraints}

Energy is also considered as an input variable or constraint in some of the analysed works. For example, Barcelo et al.\cite{002} characterised the Fog devices with their energy resources, such as power grid, battery, or energy costs, and the network links with, also, the energy costs. The limitations of devices powered with batteries are taken into account to guarantee lifetime requirements. Souza et al.\cite{064} proposed the concept of energy cells to measure the energy consumed by the underlying devices, and the optimisation took into account the number of available energy cells for the mapping of applications and devices. The objective was to minimise the excessive energy consumption in the most energy constrained devices. 
%They considered the energy as any other available resource in the devices, as is explained in the work of \cite{040}. 
%Authors in \cite{018} constrained the communication links by labelling them with the energy communication cost.

\paragraph*{Application Constraints}

Three types of application constraints have been considered for the classification of the papers in the survey: constraints related to the dependencies between the modules or services of the applications, to the workload generated over the applications, and if the users are able to define any kind of preference in the deployment of the application services.

\medskip
These constraints were not usually considered together in the papers of the survey. There were only four papers\cite{101,036,001,034} that included more than one of those application constraints, more concretely, the workload and dependency constraints. Guerrero et al.\cite{101} considered both the request rate of the applications to prioritise the placement of some application and the interrelation between the services. They considered that the interrelated services of an application should be allocated in devices within the network shortest path between the user and the Cloud provider, and the placement order was determined by the topological order of the services, placing the initial services closer to the users. Yang et al.\cite{034} analysed the workload generated from each region of the Fog domain and the dependencies between the application modules. Ottenwalder et al.\cite{036} defined the dependencies of the applications as an operator graph and also considered the load over the system to find a migration plan for the operators (services) across the devices. Arkian et al.\cite{001} considered the request rates of the applications between the Fog devices and the association between the consumers and the data. Mahmud et al.\cite{107} defined the user expectation metric that includes metrics such as the service access rate, demanded resources, and expected processing time, and this metrics is used to prioritise the placement of the applications. Additionally, the interrelation of the devices was also included in the status metric (proximity, resource availability and processing speed).

\medskip
The constraint about the dependency of the application models is the most common one. %Up to ten articles included this constraint. %In some cases, such as \cite{018}, the constraint is related to the co-location of the modules or services of an application as close as possible, probably in the same device, to reduce the network communication delays.
In the work of Skarlat et al.\cite{025}, the interactions between the application services were considered to calculate the theoretical response time of applications. Since they organised the devices in colonies, if two interrelated services are allocated in different colonies, the response time would be increased. Consequently, the optimisation algorithm should co-locate the services of an application in the same colony. Zeng et al.\cite{095} considered the interrelation between the storage (data placement) and the execution (scheduling) of the applications to minimise the application completion time by optimising the influence of the I/O time and the task completion time. Wang et al.\cite{032} proposed two algorithms which were defined by considering the interrelations between the modules of the applications, represented by a graph. The first algorithm was defined for linear-like applications and the second one for tree-like applications. Other application shapes were not considered. Rahbari et al.\cite{006} considered a symbiotic organisms search that used the relationships between the virtual machines (VM) to decide the allocation of the services on those VMs. Mahmud et al.\cite{021} presented a decentralized policy for the inter-dependent application modules that simultaneously considered the service access delay, service delivery time and device communication delays.

\medskip
User preferences were only included in few modelling efforts. Brogi and Forti\cite{011} and Brogi et al.\cite{012,102,bookchapter} created an algorithm that suggests to the user several alternatives for the deployment among a set of eligible candidates, and that permits \textit{whitelisting} those nodes that should be considered for the placement of certain services, according to user defined business policies. But they left to the users to choose how to trade-off the QoS metrics and the Fog resources consumptions, or even taking into account other types of considerations. In the work of Skarlat et al.\cite{024}, the users are allowed to define a deadline for the applications to warranty a level of QoS. Finally, Cardellini et al.\cite{038} stated that their solution could easily include user-related constraints, such as service co-location, bandwidth limitation, even tag-based constraints, but they did not implement them.

%Most proposals take into account latency etc… only a few works consider the topology of the network and blah blah 
\subsubsection{Optimised \& Comparison Metrics}
\label{sect:objective}

\begin{table}
	\centering
	\caption{\uib{Overview of the analysed articles by considered optimisation metrics.}}\label{tab:FAPPsummarymetrics}

	%Scale the table to fit on the page: or this line or the next
    \resizebox{0.5\columnwidth}{!}{
   % \scalebox{0.6}{
     \rowcolors{6}{white}{gray!20}
     \begin{tabular}{@{}|c|cc|c|c|ccc|c|@{}}\toprule
  	%\begin{tabular}{@{}cccccccccccccccccccc@{}}\toprule
%	\begin{tabular}{c|c|c|c|c|c|c|c|c|c|c|c|c|c|c|c|c|c|c|c|c|c}\toprule
    
	       &  \multicolumn{8}{c|}{Optimised \& Comparison Metric(s)} \\
    	\cmidrule{2-9}  
		
%		&& \multicolumn{6}{c|}{Infrastructure} &  Energy &  \multicolumn{3}{|c}{Application}&& \multicolumn{8}{c}{}  \\
%		\cmidrule{3-12} \cmidrule{14-21} 

	      & \multicolumn{2}{c|}{Network}&  Nodes&Energy&\multicolumn{3}{c|}{Performance} &    Cost \\
	\cmidrule{2-9}

		Ref. &

		%Optimization Objectives

		\multicolumn{1}{c}{\rotatebox[origin=c]{90}{Delay}}&
		\multicolumn{1}{c|}{\rotatebox[origin=c]{90}{Bandwidth}} &

	    \multicolumn{1}{c|}{\rotatebox[origin=c]{90}{Hardware}} &
	
	    \multicolumn{1}{c|}{\rotatebox[origin=c]{90}{Energy}} &
    
    	\multicolumn{1}{c}{\rotatebox[origin=c]{90}{QoS-assurance}}&
	 	\multicolumn{1}{c}{\rotatebox[origin=c]{90}{Execution time}}&   
	  	\multicolumn{1}{c|}{\rotatebox[origin=c]{90}{Migrations}}&
		\multicolumn{1}{c|}{\rotatebox[origin=c]{90}{Cost}} 
        
        %\multicolumn{1}{c|}{\rotatebox[origin=c]{90}{Prototype}}
        \\\midrule
%1 &2 & 3 & 4 & 5 & 6 & 7& 8 & 9 & 10 & 11 &  12 & 13 & 14 & 15 & 16 & 17 & 18 & 19 & 20& 21\\

%% SEARCH ALGORITHMS
\cite{004}\cite{bookchapterbuyya}&\checkmark&\checkmark&&\checkmark&&\checkmark&&\\
\cite{021}&&&&&\checkmark&&&\\
%18

%18
\cite{101}&\checkmark&&&&&&&\\
%20
\cite{036}&\checkmark&\checkmark&&&&&\checkmark&\\
%22
\cite{011}&&\checkmark&&&\checkmark&&&\\
%23
\cite{012}&&\checkmark&&&\checkmark&&&\\
\cite{102} \cite{bookchapter}&&\checkmark&&&\checkmark&&&\checkmark\\

\cite{109}&\checkmark&&&&&\checkmark&&\\

%29
\cite{032}&&&&&&&&\checkmark\\
%31
\cite{062}&&&&&\checkmark&&&\checkmark\\
%32
%\multirow{-11}{*}{\rotatebox[origin=c]{90}{\cellcolor{gray!10} Search }} &\cite{027}&&\checkmark&\checkmark&\checkmark&&\checkmark&&\\

%% ILP ALG
\cite{029}&\checkmark&&&&&&\checkmark&\\
%35
\cite{001}&&&&&&&&\checkmark\\
%36
\cite{034}&\checkmark&&\checkmark&&&&\checkmark&\checkmark\\
\cite{017}&\checkmark&&&&&&&\\

%38
\cite{095}&\checkmark&&&&&\checkmark&&\\
\cite{061}&\checkmark&&&&&&&\\
%42
\cite{002}&&&&\checkmark&&&&\\

\cite{107}&&&&&\checkmark&&&\\
\cite{024}&&&\checkmark&&\checkmark&&&\\
\cite{025}&&&\checkmark&&\checkmark&&&\\\

%47
\cite{066}&\checkmark&&&&&\checkmark&&\checkmark\\
\cite{019}&&&&\checkmark&&&&\\
%49
\cite{014}&\checkmark&&&\checkmark&&\checkmark&&\\

%51
\cite{030}&&&&&\checkmark&&&\\
%54-%55
\cite{105}\cite{106}&&&&&\checkmark&&&\\
%\multirow{-16}{*}{\rotatebox[origin=c]{90}{Mathematical Programming }} &\cite{038}&\checkmark&&&&&\checkmark&&\\

%% OTHER ALGORITHM
%60
\cite{033}&&&&&&\checkmark&&\\
%61
\cite{076}&\checkmark&&&&&\checkmark&&\\
\cite{053}&&&&&&&&\checkmark\\

%64
\cite{088}&&&\checkmark&&&&&\\
%66
\cite{008}&&&&\checkmark&\checkmark&\checkmark&&\checkmark\\

%67
\cite{064}&&&\checkmark&\checkmark&\checkmark&&\checkmark&\checkmark\\

%69
\cite{006}&&\checkmark&&\checkmark&\checkmark&&&\checkmark\\

%\multirow{-10}{*}{\rotatebox[origin=c]{90}{ Other Algorithms}} & \cite{110}&\checkmark& &&&&&\checkmark&\\
\bottomrule

\end{tabular}%

}
\end{table}

The ultimate objective of the FAPP is to optimise one or several metrics of the Fog domain. Usually, in a complex system such as a Fog architecture,  some of the most common optimisation objectives are contrasting, i.e., they cannot be both optimised and a trade-off between them needs to be determined. For example, if resource usage is optimised, probably the QoS of the application would be damaged.  
Some of the papers also studied the influence of their proposals in additional metrics to the ones of their optimisation metrics. By this, a general view of the effects of the FAPP proposal in the system is provided.  In this section, in addition to the optimised metrics, we have included all those metrics that have been studied, evaluated, and analysed in the results of the papers in the survey.

\begin{figure}[]
\centering
\includegraphics[width=10cm]{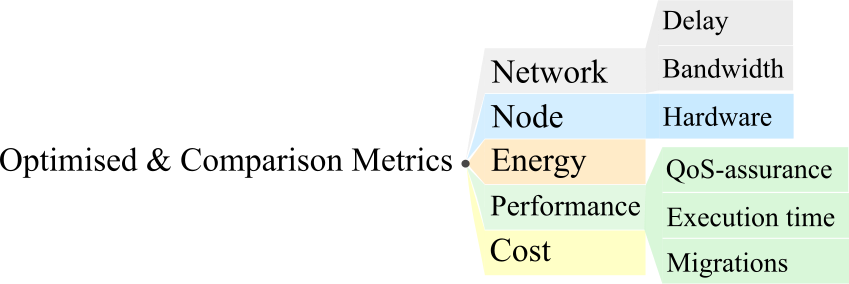}
\caption{\uib{Taxonomy of FAPP optimisation metrics.}}
\label{taxonomymetrics}
\end{figure}

\medskip
From the analysis of the papers, we defined a taxonomy of eight elements to classify the articles in terms of the considered optimisation metrics (\uib{Figure~\ref{taxonomymetrics}}). \uib{Table~\ref{tab:FAPPsummarymetrics} gives an overview of the papers classified according to this criterion}. Despite this, the papers are presented in only six groups to avoid repetition of references. In some cases, these groups are created by the union of related metrics, such as the network delay and the node execution time, that are commonly studied together. For other metrics, such as the network bandwidth and the hardware of the nodes, they resulted in not being always related to the same other metrics, and the papers that included them were already explained in other groups of metrics. If we had created a group for bandwidth or node hardware, it would have resulted in repeating the articles in several groups.

%TODO DOUBLE-CHEK: Design?, reference in text?, caption?

%Although we have classified the papers considering eight optimization metrics, the papers are organized in only six groups for a better clarity for the reader, since the groups obtained with the permutation of eight metrics would be excessive.

\paragraph*{Network Delay and Execution Time}

Probably, the most important contribution of the Fog computing paradigm is to reduce the latency of Cloud-based applications by placing them closer to the users. This latency includes the communication time and the node execution time.

\medskip
In a first set, we present the articles that included both communication and node execution time. Skarlat et al.\cite{066} obtained a decrease of the network delay and the execution time up to 39\% with regard to a baseline policy. Zeng et al.\cite{095} also considered both network and execution times. But for the latter,  they studied the computation and input/output operations separately. Cardellini et al.\cite{038} also studied other metrics, such as availability or network traffic, apart from the network delay and service time. The work proposed by Xia et al.\cite{109} minimised the application response time to improve the number of requests that are served before a fixed application deadline.
Gupta et al.\cite{004} and Mahmud et al.\cite{bookchapterbuyya} presented some baseline policies to validate their Fog simulator, and they studied the total execution time of the user requests together with the network usages and the power consumption. They compared their policies with the case of requesting services only from the Cloud provider.

\medskip
In the second set, we present the papers including the network delay but that did not consider the node execution time. There are two papers that solely considered the network delay\cite{017,061}. Souza et al.\cite{061} reduced the execution time of the application service by measuring the allocated time slots, and the results proved the reduction of the high delays of requesting the services to the Cloud provider. In some other cases, the network latency is not measured directly, and indicators such as the hop count are considered\cite{029}.
Guerrero et al.\cite{101} also minimised the hop count between the users and the placement of the services with the objective to reduce the application latency, and the network usage. The number of migrations was also included in the metrics evaluated in the experimental phase.

\medskip
Finally, the number of papers considering the execution time but without including the network delay is very reduced. For example, Wen et al.\cite{033} solely considered the execution time, showing improvement around the 30\%. Additionally, Mennes et al.\cite{076} had the objective of maximising the number of applications deployed on Fog devices, but they also measured the execution time of the application in the results of their experiments.

\paragraph*{QoS-assurance}

Some approaches, instead of minimising network or execution times, aimed at increasing the Quality of Service (QoS) satisfaction. QoS is directly related to those times, but its improvement does not necessarily result in a reduction of the times. For example, the QoS can be measured as the percentage of requests that are executed before a time deadline. The improvement of the QoS involves then to keep the execution times below this threshold, but the minimisation of the times is not required.

\medskip
Brogi and Forti\cite{011} and Brogi et al.\cite{012,102,bookchapter} studied the QoS in terms of latency and network bandwidth. They also considered resource consumption of the Fog devices and they proposed a novel cost model for Fog devices.
In the work of Mahmud et al.\cite{021}, the objective of the optimisation was to reduce the number of active Fog devices. But this optimisation was constrained with the warranty of satisfying the QoS level, i.e., shorter execution times than the application deadlines. Consequently, results about the percentage of deadline satisfaction were presented, showing important improvements.
Skarlat et al.\cite{024,025} maximised the resource usage of the Fog devices to maximise the number of applications deployed on the Fog layer, while the latency and QoS are not damaged. The results showed that the Fog landscape was used for the 70\% of the services, reducing  30\%  of the execution cost, without affecting the QoS. 

\medskip
Venticinque and Amato \cite{030} considered the QoS in terms of the number of request and transactions processed per unit of time. The authors also included the results of the execution time and the resource usages of the devices.
Cardellini et al.\cite{105,106} also studied the QoS measured with data obtained from each node, such as utilisation, availability, and network metrics. All the nodes are informed of the QoS of other nodes with the use of a gossip-based dissemination schema. The authors presented the experiment results in terms of application availability, application latency, network traffic and node utilization.

\medskip
Mahmud et al.\cite{107} defined three metrics to study the QoS of their proposal: network relaxation ratio, processing time reduction ratio and resource gain. Additionally, they also presented results about deadlines, costs and packet losses.
Finally, Zhang et al.\cite{088} optimised the QoS by providing a suitable utilization of the nodes. The algorithm ensures an optimal amount of hardware resources for the allocated applications.

\paragraph*{Migrations}

An important consequence of bringing the applications near to the users is the increase of the network traffic due to the application migrations. Consequently, some of the studies have dealt with the minimization of the number of migrations or their effects on the system.

\medskip
Apart from the already cited work of Velasquez et al.\cite{029}, where the number of migrations was minimised along with the network delay, migrations were also optimised by Ottenwalder et al.\cite{036}, by minimising the network utilization without damaging the network latency, and by Yang et al.\cite{034}, where the migration was reduced along with the latency, the resource usage, and the provider cost.

\medskip
Finally, Filiposka et al.\cite{110} studied the cost of migrations to just migrate the applications if the obtained benefits were greater than the overhead generated in the system by the movement of the application between nodes. They also studied the network latency in terms of hop counts.

\paragraph*{Cost}

Cost is a common minimisation objective in resource management proposals in computation-as-a-service domains. In Fog domain the evaluation of the cost is still in a very initial phase, but there are some preliminary works dealing with its optimisation.

\medskip
In addition to the already cited works\cite{102,bookchapter,066}, there are other four papers that included the cost in their evaluation and optimisation. Zhang et al.\cite{053} studied the cost of the transmission delay and the service execution.
Arkian et al.\cite{001} optimised the overall cost of the deployment of the applications, while the QoS is guaranteed, by allocating them in the devices with the smallest costs. They additionally studied the power consumption and the service latency. Wang et al.\cite{032} addressed the minimisation of the cost of each physical node and link, ensuring that the devices would not be overloaded.

\medskip
Finally, Hong et al.\cite{062} optimised the cost of the application deployment by selection of the provider from a federated pool of Cloud providers. Additionally, other three objectives were also minimised: the available resources on the devices, the network distance between the users and the services, and between the nodes which allocate interrelated services.

\paragraph*{Energy}

The energy and the power consumption are also one of the most important concerns among the authors of the analysed papers. The energy optimisation has been addressed from different point of views. For example, Barcelo et al.\cite{002} defined a linear characterised function of the energy cost, and they minimised it. Their proposal was able to reduce the overall power consumption by more than an 80\%.

\medskip
Huang et al.\cite{019} were more focused on the reduction of the communication energy cost, by placing in the same device interrelated services and, consequently, reducing the number of communications between devices and their hop count. The experiments showed an improvement of 10\% energy savings.
In other cases, the energy was optimised along with other metrics: (a) the trade-off between energy consumption and end-user delay\cite{014}, (b) optimisation of the energy consumption with the network usage and the execution cost\cite{006}, resulting in improvements of 18\% for the energy consumption, 1.17\% for the network usage and 15\% for the execution cost, (c) balancing the energy consumption and the resource usage in the Fog devices\cite{064}, (d) or reducing the application execution time, the power consumption of the devices, and the cost of the services migrations\cite{008}. 

\medskip
In a few cases, the energy was not the optimisation objective, but it was analysed to validate the benefits of the proposals. For example, Taneja et al.\cite{027} mainly addressed the minimisation of the application execution time, but they also analysed other common metrics such as network usage and energy consumption.

\section{Conclusions: Open Problems and Research Challenges}
\label{openchallenges}

In this section, we conclude by pointing to some open challenges and future directions that can be explored to better approach FAPP.

\medskip
First of all, whilst search and mathematical programming have been thoroughly investigated in the literature, more modern techniques are to be applied to FAPP. Particularly, based on the first promising results they obtained on FAPP, it would be interesting to study further the applicability of other genetic or evolutionary algorithms, swarm optimisation techniques, deep learning, network science and game theoretical approaches to FAPP. 
Additionally, very few approaches available nowadays are distributed, whilst those solutions might scale better and show stronger resilience in highly dynamic Fog infrastructures. In these regards, devising decentralised algorithms to be applied to solve FAPP online and without relying on control nodes would be important and crucial to the success of the Fog.

\medskip
Naturally, whilst exploring new proposals, it is important to compare them to the related state-of-the-art techniques with respect to the execution time (utterly important in Fog scenarios) and -- possibly -- to the achieved results (in terms of a standard set of common metrics). To this end, the design of a set of benchmark examples (based on established standards like TOSCA Yaml \cite{tosca}) would make it possible to systematically compare and contrast different approaches to FAPP as well as to quantify performance improvements or degradation. Such examples should possibly consider all the attributes that have been modelled in the literature (e.g., hardware, software, IoT, QoS, energy) and should come with an optimal candidate solution, determined by exhaustive techniques. Another possibility towards this direction is to exploit the solid theory of complex networks both to generate test topologies and to analyse the obtained results. %Finally, extended availability of open-source codebases and prototypes would definitely ease this tasks.

\medskip
From a modelling perspective, none of the surveyed studies considered security aspects when determining optimal application placements. Security will play a crucial role in the success of the Fog paradigm and it represents a concern that should be addressed \textit{by-design} at all architectural levels \cite{openfogarch}. Therefore, there is a clear need to (quantitatively) evaluate whether an application will have its security requirements fulfilled by the (Cloud and Fog) nodes chosen for the deployment of its components. Furthermore, due to the mission-critical nature of many Fog applications (e.g., e-health, disaster recovery), it is important that the techniques employed to perform security analyses to solve FAPP are well-founded and, possibly, explainable. 

\medskip
Similarly, mobility of Fog and IoT nodes was considered in very few of the reviewed works, even though it is a parameter that cannot be neglected in Fog scenarios. Indeed, many Fog verticals (e.g., autonomous vehicles, flying drones) include nodes that move, and that continuously and opportunistically connect/disconnect from other devices. In general, few authors considered infrastructure variations due for instance to changing topologies (i.e., available nodes), network traffic (i.e., latency, bandwidth), or workload conditions (i.e., available node hardware). Future research in the field of FAPP should, therefore, devise and tweak novel models that account for these typical traits of Fog computing, so to understand how application placement can be adaptively adjusted to this phenomenon. 

\medskip
Analogously, Fog computing exists in continuity with both the IoT and the Cloud. Hence, more effort should be made to consider the integration and simultaneous management of these three entities which was neglected in many works. Particularly, QoS attributes that define the reachability of IoT and Cloud nodes from different Fogs showed to be key in leading the search towards better placements. In line with this effort of considering the Cloud to Things continuum as a unique system, the possibility of some application components to be deployed in different flavours depending on the resources of target deployment nodes (like in \textit{Osmotic Computing} \cite{osmotic}) is to be studied yet. Overall, understanding how the proposed methodologies could be exploited to work with production-ready tools for Fog application management (e.g., CISCO FogDirector \cite{fogdirmime}) would be surely of interest.

\medskip
To conclude, most of the experiments were carried out in small to medium scale simulated environments, often without disclosing the codebase to repeat them. Future research in this field should prototype more versatile and well-documented simulators that can be used to experiment with different strategies or algorithms and that permit evaluating proposals over large-scale, lifelike, examples. Last but not least, real Fog testbeds could be realised with the help of those industries that are currently shaping Fog computing, so to actually test and assess the proposed solution strategies.

\bibliography{sample}

\end{document}